\documentstyle[10pt]{espart}
\begin{document}
\begin{frontmatter}
\title{The Adsorption of H$_2$O on TiO$_2$ and SnO$_2$~(110) \\
Studied by First-Principles Calculations}
\author{J. Goniakowski and M. J. Gillan}
\address{Physics Department, Keele University, Staffordshire ST5 5BG, U.K.}
\date{\today}
\begin{abstract}
First-principles calculations based on density functional theory
and the pseudopotential method have been used to investigate the
energetics of H$_2$O adsorption on the (110) surface of TiO$_2$ and
SnO$_2$. Full relaxation of all atomic positions is performed on
slab systems with periodic boundary conditions, and the cases of full
and half coverage are studied. Both molecular and dissociative
(H$_2$O $\rightarrow$ OH$^-$ + H$^+$) adsorption are treated, and
allowance is made for relaxation of the adsorbed species to unsymmetrical
configurations. It is found that for both TiO$_2$ and SnO$_2$
an unsymmetrical dissociated configuration is the most stable. The
symmetrical molecularly adsorbed configuration is unstable with
respect to lowering of symmetry, and is separated from the fully
dissociated configuration by at most a very
small energy barrier. The calculated dissociative adsorption energies for
TiO$_2$ and SnO$_2$ are in reasonable agreement with the results
of thermal desorption experiments. Calculated total and local
electronic densities of states for dissociatively and molecularly
adsorbed configurations are presented and their relation with
experimental UPS spectra is discussed.
\end{abstract}
\end{frontmatter}

\section{Introduction}
Most oxide surfaces react readily with water and become partially
covered with molecular H$_2$O or hydroxyl groups on exposure to
air under ambient conditions. This adsorption of water has important
consequences for surface processes such as catalysis and gas sensing.
It is widely accepted that molecular water will generally bind to
most oxide surfaces through the attraction of its electric dipole
to the ionic charges. However, the mechanisms by which dissociative
adsorption occurs are still very poorly understood. It has sometimes
been suggested that surface defects such as vacancies or steps are
essential to water dissociation, and there is no doubt that there are
some oxide surfaces for which this is true. For example, recent
calculations using both semi-empirical~\cite{gon95} and
first-principles~\cite{lan94,sca94}
techniques have shown that water dissociation is not energetically
favorable on the perfect MgO~(100) surface, but that it is
favorable at steps and corners. However, it is unlikely that
defects are essential for dissociative adsorption in general,
since oxide surfaces vary enormously
in their geometry and electronic structure.

The aim of this paper is to present first principles calculations
on the energetics of H$_2$O adsorption on the SnO$_2$ and
TiO$_2$~(100) surfaces. Both materials have the rutile crystal structure,
and the (110) surface is the most stable when the materials are
stoichiometric. The stoichiometric (110) surface is far from flat,
so we are dealing with a situation which is geometrically
very different from the MgO~(100) surface. Although they have the
same geometry, SnO$_2$ and TiO$_2$ are electronically different,
since Ti is a transition element and Sn is a group~IV element.
Our hope is therefore that by comparing with previous work on the
interaction of H$_2$O with MgO~(100) we can learn about the
influence both of geometry and of electronic structure.
Our choice of SnO$_2$ and TiO$_2$ is also motivated by their
technological importance. SnO$_2$ is widely used in gas-sensing
devices, and adsorbed water is known to have an important
effect on the adsorption of other molecules~\cite{tan89}. The interaction of
TiO$_2$ with water is important in photocatalytic applications~\cite{fuj72}.

A considerable experimental effort has gone into the study of H$_2$O
adsorption on TiO$_2$~(110), though a clear picture has yet to emerge.
Hydroxyl groups on the surface have been detected at 300~K in
ultraviolet photoelectron spectroscopy (UPS)
experiments by Henrich {\em et al.}~\cite{hen77}, and their
presence was later confirmed in synchrotron radiation studies
by Madey's group~\cite{kur89,pan92}. The amount of  dissociated water
was less than a monolayer, and was weakly dependent on the oxygen
vacancy concentration before adsorption. The minor influence of
point defects and steps on the reactivity with water has also been
reported by Muryn {\em et al.}~\cite{mur91}. Experiments using
a combination of thermal desorption spectroscopy (TDS), x-ray
photoelectron spectroscopy (XPS) and work-function measurements
have recently been reported~\cite{hug94}. A three-peak desorption
spectrum was found,
with features at 170 and $\sim$275~K and a high-temperature tail
up to $\sim$375~K. The two low-temperature features appear to be
due to molecular water and the high-temperature tail to the reaction
of surface OH groups to form H$_2$O.
A widely accepted model is that of Kurtz {\em et al.}~\cite{kur89}, who
proposed
that adsorption on nearly perfect TiO$_2$~(110) occurs by molecular adsorption
of H$_2$O at 5-fold coordinated cation sites, followed by dissociation
to give OH$^-$ attached by its oxygen end to the cation, and H$^+$ bonded
to lattice oxygen to form a second type of hydroxyl group.
The presence of
at least two types of adsorbed OH$^-$ has been indicated by infrared
spectroscopy~\cite{mun71,jon71}.

The experimental situation for SnO$_2$ is rather similar, although
the experimental evidence is less extensive. Recent
experiments by Gercher and Cox~\cite{ger95} using a combination of
TDS and UPS show the presence of dissociatively adsorbed water
on SnO$_2$~(110). On the perfect stoichiometric
surface, three distinct TDS peaks are seen at $T$ = 200,
300 and 435~K. The 200~K peak is attributed entirely and the 300~K peak
mainly to molecular H$_2$O. The feature at 435~K is assigned to
disproportionation of surface OH$^-$ to form H$_2$O. As in the
case of TiO$_2$, surface defects do not appear to play a major role.

Few theoretical investigations of H$_2$O adsorption on rutile
surfaces have been reported. Jaycock and Waldsax~\cite{jay74} performed
calculations for the TiO$_2$ (110) and (100) surfaces using
an empirical interaction model, which predicted adsorption energies much
greater
than the experimental values. The dissociative adsorption of single molecules
on clusters embedded in a point-charge lattice has been studied using
the $X \alpha$ approach for TiO$_2$~(110) by Tsukada
{\em et al.}~\cite{tsu83}. The dissociative adsorption of single water
molecules
on TiO$_2$ as well as the case of full hydroxylation
have been studied using a semi-empirical technique by
Goniakowski {\em et al.}~\cite{gon93} and Goniakowski and Noguera~\cite{gon95}.
Both molecular and dissociative adsorption of water on TiO$_2$ have been
treated using a periodic Hartree-Fock approach by Fahmi and
Minot~\cite{fah94}, who find a strong preference of water to dissociate.
The above studies considered only very symmetrical
configurations of adsorbed species, and all found that both
molecular and dissociative adsorption are energetically favorable.
The only attempt to study relaxation of surface species to unsymmetrical
configurations has been the very recent semi-empirical study by
Bredow and Jug on the adsorption of H$_2$O on rutile and anatase TiO$_2$
surfaces~\cite{bre95}. To our knowledge no theoretical attempts to model
water adsorption on SnO$_2$ surfaces have been reported.

The present work is based on density functional theory (DFT) in the
pseudopotential approximation. This approach has been widely
used for oxides, including
MgO~\cite{vit92}, Li$_2$O~\cite{Ian92}, Al$_2$O$_3$~\cite{Ian93},
TiO$_2$~\cite{ram94a,ram94b}
and SnO$_2$~\cite{IanYY}, and is known to give accurate results
for the energetics of perfect crystals, lattice defects,
surfaces and molecular adsorption. We have recently reported a
detailed study of the stoichiometric and reduced SnO$_2$~(110)
surface using this approach~\cite{IanXX}.
However, since the energetics of dissociation is not generally
described very accurately within the standard local density
approximation (LDA), we include gradient corrections in the present work,
using the Becke-Perdew scheme~\cite{bec88,per86}.
We present here results
on the energetics and relaxed geometry of both molecularly
and dissociatively adsorbed water on SnO$_2$ and TiO$_2$~(110),
and we study both symmetrical and unsymmetrical modes of adsorption.
In addition, we report total and local electronic densities of
states, through which we attempt to make contact with recent
spectroscopic measurements.

\section{Techniques}
\subsection{General background}
The general principles of DFT and the pseudopotential method have
been described in the literature~\cite{hoh64,koh65,jon89,gil91,pay92}.
Within the pseudopotential approximation only
valence electrons are represented explicitly in the calculations, the
valence-core interaction being represented by non--local pseudopotentials
which are generated by first principles calculations on isolated atoms.
The calculations are performed using periodically repeating geometry, the
occupied orbitals being expanded in
a plane wave basis. This expansion includes all plane waves whose kinetic
energy $E_k = \hbar^2k^2/2m$ ($k$ the wavevector, $m$ the electron mass)
is less than the cut-off energy $E_{\rm cut}$, chosen so as to ensure
convergence with respect to the basis set.

In the present work the self-consistent ground state of the system was
determined using a band-by-band conjugate gradient technique to minimize
the total energy of the system with respect to the plane-wave coefficients.
Equilibrium positions of ions were determined by a steepest descent
method.  The calculations were performed using the CETEP code~\cite{cla92}
(the parallel version of the serial code CASTEP~\cite{pay92}), running on
the 64-node Intel iPCS/860 machine at Daresbury Laboratory.

\subsection{Generalized Gradient Corrections}
Standard DFT calculations make use of the LDA, which treats the
electron density as locally uniform. Although this is highly
successful for many purposes, it does not give accurate results
for energy differences involving changes of bonding, which are of
interest here. In the last few years, methods have been developed
for improving the LDA through `generalized gradient' approximations
(GGA). It has been shown that these corrections lead to improvement
in the calculation of total energies of atoms and
molecules~\cite{per86,lan83,lan85,pw86,kut88,bos90,mly91},
cohesive energies~\cite{gar92,kon90,ort92}, and the energetics of
molecular adsorption and dissociation
on metal surfaces~\cite{whi94a,hu94,phi94,gun94}.
We have recently reported a comparison
of the influence of two popular GGA schemes -- those due to
Perdew and Wang~\cite{per86,pw86} and Becke and Perdew~\cite{bec88,per86}) --
on the bulk and surface properties of TiO$_2$ and SnO$_2$~\cite{unp}. We found
that the Becke-Perdew method gives better equilibrium lattice
parameters, while giving essentially the same surface properties as
Perdew-Wang, and we employ Becke-Perdew in the present work. We use
the technique of White and Bird~\cite{whi94} for incorporating
GGA within the DFT-pseudopotential calculation. In some places,
we shall also report LDA results for comparison; these were obtained
using the Ceperley-Alder (CA) exchange-correlation function~\cite{cep80}.

\subsection{Generation of pseudopotentials}
First-principles, norm-conserving pseudopotentials in Kleinman-Bylander
representation~\cite{kle82} were generated using the
optimization scheme of Lin {\em et al.}~\cite{lin93} in order to
reduce the required value of the plane-wave cut-off $E_{\rm cut}$.
The pseudopotentials used in GGA calculations were constructed in a
consistent way by including GGA in the generation scheme.
The Sn
pseudopotential was generated using the $5s^25p^2$ configuration for
$s$- and $p$-wave components, and the $5s^15p^{0.5}5d^{0.5}$
configuration for the $d$-wave. The core radii were equal to 2.1, 2.1 and
2.5~a.u. for the $s$, $p$ and $d$ components respectively.
The Ti pseudopotential was generated using
the $4s^{1.85}3d^2$ configuration for the $s$ and $d$ waves
and the $4s^14p^{0.5}3d^{0.5}$ configuration for the
$p$ wave, with core radii of 2.2, 1.5 and 2.4~a.u. for $s$, $p$ and $d$ waves
respectively.
The oxygen pseudopotential used in our LDA calculations
was generated
using the $2s^22p^4$ configuration for the $s$ and $p$ waves and the
$2s^22p^{2.5}3d^{0.5}$ configuration for the $d$ wave,
with a single core radius of 1.65~a.u. For the
gradient-corrected oxygen pseudopotential, we have used the single
configuration $2s^22p^{3.5}3d^{0.45}$ and the same core radius.
The use of a core radius of 1.65~a.u. means that there is an appreciable
overlap of the oxygen and metal core spheres in the SnO$_2$ and TiO$_2$
crystals, and in principle this could cause inaccuracies. However,
direct comparisons of the present results with our earlier work
on SnO$_2$~\cite{IanXX}, which employed an oxygen pseudopotential with the
smaller core radius of 1.25~a.u., show that any errors due to core overlap are
very small.

The calculations have been done using a plane wave cut-off of 600~eV for
SnO$_2$ and 1000~eV for TiO$_2$. Our tests show that with these cut-offs
the energy per unit cell is converged to within 0.2~eV.

\subsection{Densities of States}
Electronic densities of states (DOS) associated with the ground state were
calculated using the tetrahedron method~\cite{jep71,leh72}, with $k$-point
sampling corresponding to 750 tetrahedra in the whole
Brillouin zone. In addition, local densities of states (LDOS)
were calculated by taking contributions
only from chosen regions of real space -- in practice
we have used spheres (radius 1.5~\AA) centered on chosen
surface oxygen atoms. For presentation purposes,
we have broadened the calculated DOS and LDOS by Gaussians of width 0.5~eV.
In the ground state calculations the Brillouin zone sampling is performed
using the lowest order Monkhorst-Pack set of $k$ points~\cite{mon76},
as in our previous work on SnO$_2$~\cite{IanYY,IanXX}.

\section{Tests on SnO$_2$, TiO$_2$ and H$_2$O}

\subsection{Perfect crystals and (110) surfaces}
A more detailed discussion of the influence of gradient
corrections on the calculated parameters of the SnO$_2$
and TiO$_2$ perfect crystals and (110) surfaces is given in a
separate paper~\cite{unp}. Here we present the system used in the
calculations and recall the principal results obtained
using the Becke-Perdew GGA scheme.

\begin{figure}
\caption{ The unit cell of the rutile crystal structure. Cations are
represented by small black circles and oxygen atoms by large white circles.}
\end{figure}
The six-atom rutile unit cell of SnO$_2$ and TiO$_2$ is shown in Fig.~1.
The equilibrium structure has been determined by relaxation with
respect to the lattice parameters $a$ and $c$ and the
internal parameter $u$. The equilibrium values of these
parameters calculated using the Becke-Perdew form of GGA are given in Table~1.
\begin{table}
\caption{Comparison of theoretical and experimental \protect \cite{wyckoff}
values of lattice parameters $a$ and $c$ and the internal coordinate $u$ of
SnO$_2$ and TiO$_2$. The theoretical values are calculated using
the Becke-Perdew form of GGA.}

\begin{tabular}{lcccc}
\hline
\hline
&\multicolumn{2}{c}{SnO$_2$}&\multicolumn{2}{c}{TiO$_2$} \\
& calc.	& expt.	& calc.	& expt. \\ \hline
$a$ (\AA)&  4.809 & 4.737 & 4.747 & 4.594 \\
$c$ (\AA)&  3.159 & 3.186 & 3.039 & 2.958 \\
$c/a$ (\AA)&  0.657 & 0.673 & 0.640 & 0.644 \\
$u$ (\AA)&  0.307 & 0.307 & 0.305 & 0.305 \\
\hline
\hline
\end{tabular}
\end{table}
The agreement of $a$ and $c$ with experiment is very
satisfactory for SnO$_2$ and  acceptable for TiO$_2$;
the values of $u$ are excellent in all cases.

Our calculations on the stoichiometric (110) surface of
the materials have been done with the usual repeating slab geometry.
The rutile structure can be regarded as consisting of (110) planes of atoms
containing both metal (M) and oxygen (O) atoms, separated by planes
containing oxygen alone, so that the sequence of planes is
O - M$_2$O$_2$ - O - O - M$_2$O$_2$ - O etc. The entire crystal can
then be built up of the symmetrical 3-plane O - M$_2$O$_2$ - O units.
The slabs we use contain three of these units, and our repeating cell
contains 18 atoms (6 M and 12 O). The perfect (110) surface
consists of rows of bridging oxygens lying above a metal-oxygen
layer (see Fig.~2).
\begin{figure}
\caption{ Atomic structure of the clean (110) surface. 6- and 5-fold
coordinated cations are represented by small black circles and small black
circles with white centers respectively. In plane oxygen atoms
are represented by large white circles and bridging oxygens
by large gray circles.}
\end{figure}
The vacuum separating the slabs has been taken wide enough to
ensure that interactions between neighboring slabs are small
even when adsorbed H$_2$O and OH$^-$ are present.
The width we use corresponds to two O - M$_2$O$_2$ - O units, and
is such that planes of bridging oxygens on
the surfaces facing each other across the vacuum are
separated by about 6.8~\AA. This vacuum
width is somewhat greater then was used in our earlier
calculations on SnO$_2$~(110)~\cite{IanXX}.
(It will be convenient in the following
to specify slab thickness and vacuum width in terms of the equivalent
width of O - M$_2$O$_2$ - O units.)

The surface structure has been determined by relaxing the
entire system to equilibrium. As in our previous work on SnO$_2$~(110),
and the work of
Ramamoorthy {\em et al.} on TiO$_2$~(110)~\cite{ram94a}, we find
displacements of the surface atoms of order 0.1~\AA, with
5-fold and 6-fold coordinated metal atoms moving respectively into and
out of the surface, in-plane oxygens moving out and bridging
oxygens moving very little. The modifications of the bond lengths
between the surface atoms with respect to the perfect crystal for
Becke-Perdew GGA calculations are given in Table~2.
\begin{table}
\caption{Calculated bond length modifications on SnO$_2$~(110)
and TiO$_2$~(110) with respect to the bulk values for Becke-Perdew
GGA exchange-correlation. For atom indexing see Fig.~2}
\begin{tabular}{lcc}
\hline
\hline
&SnO$_2$&TiO$_2$  \\  \hline
O$_{\rm I}$ -- M$_{\rm I}$ &  -3.8\% &  -5.5\%  \\
O$_{\rm IV}$ -- M$_{\rm II}$ &  -4.2\% &  -5.6\%  \\
O$_{\rm II}$ -- M$_{\rm II}$ &  -1.2\% &  -1.2\%  \\
O$_{\rm II}$ -- M$_{\rm I}$ &   2.9\% &   2.8\%  \\
O$_{\rm III}$ -- M$_{\rm I}$ &   4.8\% &   4.5\%  \\
\hline
\hline
\end{tabular}
\end{table}

We find that the relaxed surface energy of SnO$_2$~(110) is
1.16~Jm$^{-2}$ and 0.84~Jm$^{-2}$ for TiO$_2$~(110). These energies are,
as already disscused in Ref.~\cite{unp}, substantially lower than the
corresponding LDA results.

The electronic DOS and the LDOS on bridging oxygen calculated for
the valence band of the slab systems using the Becke-Perdew
form of GGA are shown in Fig.~3.
\begin{figure}
\caption{Calculated valence band DOS for
clean (110) surfaces of TiO$_2$ and SnO$_2$. The dashed line
represents the LDOS on the bridging oxygen site O$_I$. For the sake
of presentation the local contribution has been scaled by a factor of 5.}
\end{figure}
The main features of the calculated surface DOS of SnO$_2$
have already been discussed in our previous paper~\cite{IanXX}. We note
particularly the splitting off of a narrow band of states at
the top of the O(2s) band and the appearance of a
sharp peak at the top of the O(2p) valence band. Both these
features are associated with bridging oxygens. The features are
also present for TiO$_2$, although the splitting is less marked.

\subsection{H$_2$O and OH$^-$ molecules}
We have calculated the equilibrium geometry and electronic structure
of the H$_2$O and OH$^-$ molecules. Since our techniques require us
to use periodic boundary conditions, the calculations are actually
performed on periodic arrays of molecules, with the repeating cell
chosen to be a cube of side $R$. For H$_2$O, we find that the choice
$R$ = 7~\AA\ is large enough to render the interactions between periodic
images negligible. Matters are not so simple for the OH$^-$ molecule,
since it carries a net charge. To make mathematical sense of calculations
in which the repeating cell is charged, it is essential to introduce
a uniform compensating background, whose charge density is chosen so that
the net charge in the unit cell vanishes. This technique is well
established, and has often been discussed in the literature. However,
as discussed by Leslie and Gillan~\cite{les85},
the total energy still converges very
slowly with increasing cell size, and the leading term in its deviation
from the asymptotic value is $- \alpha / R$, where $\alpha$ is the
Madelung constant for the appropriate periodic array of point charges.
This leading
correction can therefore be subtracted exactly, and we are left with
a total energy which converges reasonably quickly to the value for the
molecule in free space. In our calculations on OH$^-$, we find that
the corrected total energy is converged to
better than 0.1~eV for $R$ = 13~\AA,
and the equilibrium bond length is converged well before this. Comparison
of the calculated and experimental structural parameters of
H$_2$O and OH$^-$ (Table~3) confirms that the molecules are described
accurately by the present methods.
\begin{table}
\caption{Comparison of theoretical and experimental values of bond
lengths d$_{O-H}$ and bond angle
$\protect \angle $H-O-H of H$_2$O and OH$^-$.
The theoretical values are calculated using the Ceperley-Alder
form of LDA~(CA), and the Becke-Perdew form of GGA~(BP).
Experimental values are taken from
Refs. \protect \cite{eisenberg,herzberg}}
\begin{tabular}{llcccccc}
\hline
\hline
&& \multicolumn{2}{c}{$d_{O-H}$ (\AA)}&\multicolumn{2}{c}{$\angle H-O-H$}  \\
\hline
H$_2$O  & CA & 0.977 &  (2.1\%) & 104.8 & ( 0.3\%)  \\
        & BP & 0.975 &  (1.9\%) & 103.6 & (-0.9\%)  \\
      & expt.& 0.957 &          & 104.5 &           \\ \hline
OH$^-$  & CA & 0.980 &  (1.0\%) &  &   \\
        & BP & 0.979 &  (0.9\%) &  &   \\
       &expt.& 0.970 &          &  &   \\
\hline
\hline
\end{tabular}
\end{table}

Since we are concerned with dissociation of H$_2$O in this paper, we have
tested the influence of the GGA on its dissociation energy. The reaction
of interest is H$_2$O $\rightarrow$ OH$^-$ + H$^+$. To compare
our results with experimental data, we start from the experimental
dissociation energy for the reaction H$_2$O $\rightarrow$ OH + H, which
is 5.11~eV. We add to this the ionization energy of H (13.60~eV) and
we subtract the electron affinity of OH (1.83~eV), to obtain 16.88~eV.
Finally, addition of the zero-point vibrational energy
of H$_2$O (0.57~eV) and subtraction
of the corresponding quantity for OH (0.23~eV) gives us the
value 17.22~eV which can be compared with the calculations. The
calculated values with LDA and GGA are 16.6 and 16.7~eV, so that
even with gradient corrections there is a residual error of $\sim$~0.5~eV.

In order to compare with the electronic DOS reported later,
it is useful to note that the single-particle energies of the occupied
molecular orbitals (MO) of H$_2$O can be related to experimental measurements.
As usual, one must be cautious about comparing Kohn-Sham single particle
energies with spectroscopic energies, but it is known empirically that for
occupied states the comparison is usually justified. We therefore compare
the differences of our calculated MO energies with the corresponding
differences of measured ionization energies.
In the usual notation, the MO states of H$_2$O are
$2 a_1$, $1 b_2$, $3 a_1$ and $1 b_1$.
Our calculated separations of $2a_1 - 1b_2$,
$1b_2 - 3a_1$ and $3a_1 - 1b_1$ (experimental values from ref.~\cite{bal78}
in parentheses)
are 11.96~eV (13.6~eV), 3.64~eV (3.8~eV) and 2.12~eV (2.0~eV) respectively.
Our calculated
separations of $2\sigma - 3\sigma$ and $3\sigma - 1\pi$ levels in OH$^-$
are respectively 12.2~eV and 3.5~eV.

\section{Surface adsorption of H$_2$O}
We turn now to our calculations on dissociative and molecular
adsorption of H$_2$O on SnO$_2$ and TiO$_2$ (110). The calculations
are done using the same repeating-slab geometry used for the
bare surface, and we need to pay attention to the effects of slab
thickness and vacuum width. Another important technical
question is whether it is better to perform the calculations
with the particles adsorbed only on one surface of each slab
or on both surfaces. We refer to these as the one-sided and two-sided
geometries. There is a strong argument for working with
the symmetrical two-sided geometry in which the same particles are adsorbed on
opposite surfaces, because any possible dipole moment of the
repeated cell is then eliminated, and convergence of the adsorption
energy with increasing slab thickness is likely to be improved.
We have made tests which confirm that this is the case, and all our
calculations have therefore been made using the two-sided
geometry.

Our tests on the effects of slab thickness and vacuum
width were performed on the SnO$_2$ system in which
H$^+$ is adsorbed on top of every bridging oxygen and OH$^-$ is
adsorbed at every 5-fold Sn site (see Fig.~4a).
\begin{figure}
\caption{ Atomic structure of fully hydroxylated (110) surface of rutile
for a) SD, b) UD and c) SM adsorption geometries.
Atom symbols as in Fig.~2, with adsorbed oxygen atoms
represented by large white circles and hydrogens by small white circles.}
\end{figure}
In these tests, every
atom in the system is relaxed to its equilibrium position. Using
the two-sided slab geometry mentioned above, we find that increase
of the vacuum width from two to three O - Sn$_2$O$_2$ - O units changes the
adsorption energy by only 0.01~eV per water molecule, and increase of
the slab thickness from three to four units gives a change of 0.03~eV
per molecule. If the one-sided slab geometry is used, the changes
are nearly ten times as great. The results to be presented
have all been obtained with a slab thickness of three units and
a vacuum width of two units. We stress that in all the calculations
that follow the entire system is fully relaxed to equilibrium.

As noted in the Introduction, we cannot be sure in advance how H$_2$O
will prefer to adsorb. To study dissociative adsorption, we begin with
the symmetrical full-coverage case mentioned above (see Fig.~4a).
We then study symmetry-lowering distortions from this configuration.
A similar strategy is followed for the case of molecular adsorption.
The effect of going to lower coverage is than briefly examined for the
dissociative case.

\subsection{Symmetrical dissociative adsorption (SD)}
The dissociative adsorption energy is obtained from the fully relaxed total
energy of the slab system in which H$^+$ and OH$^-$ are adsorbed at both
surfaces. It is calculated in the natural way by subtracting
this energy per unit cell from the corresponding energy for the
bare slab plus the energy of a pair of H$_2$O molecules; the result is,
of course, divided by two, since two H$_2$O molecules (one on each
surface) are adsorbed per unit cell. As usual,
a positive adsorption energy means that the total energy decreases
when the molecule is adsorbed. In the symmetrical dissociative (SD)
case, our LDA calculations for SnO$_2$ and TiO$_2$ (110) yield
adsorption energies of 1.19 and 0.91~eV per H$_2$O respectively.
Inclusion of gradient corrections lowers the adsorption energies
considerably to 0.48 and 0.45~eV per H$_2$O. It is clear from this
that gradient corrections have an extremely important effect on the
calculated adsorption energies. A similar effect of lowering of the adsorption
energy by inclusion of GGA has already been reported for adsorption on metal
surfaces~\cite{hu94,phi94}, leading to a substantial
improvement with respect to experimental results.

We find that the dissociative adsorption of water causes the equilibrium
surface
structure to change significantly. The relaxations found for the clean surface
are greatly reduced and in some cases reversed. The adsorption causes
5- and 6-fold cations to
move respectively out of and into the surface, the in-plane oxygens
to move outward, and the bridging oxygens to move slightly outward, relative to
the relaxed clean surface.
Compared with free hydroxyl molecules, surface O--H bonds are
considerably shorter, the shortening being  especially pronounced for
the OH group attached to the 5-fold surface cation. Similarly,
the bond between the OH group and the surface cation is
much shorter than the bulk O--cation bond, resembling rather the
inter-atomic distance in the corresponding diatomic molecule.
Details of inter-atomic bond length changes for
both materials (with respect their perfect crystal values), with and without
gradient corrections, are presented in Table~4.
The changes
of O--H bond lengths are given with respect to the free OH$^-$ molecule.
It is clear from the table that, in contrast to the adsorption energies,
the relaxed structure of the hydroxylated surface is little
affected by gradient corrections.
\begin{table}
\caption{Calculated bond length modifications on hydroxylated SnO$_2$~(110)
and TiO$_2$~(110) with respect to the bulk values for LDA (CA), and GGA (BP)
forms of exchange-correlation. The modifications of O--H bond lengths are
given with respect to the free OH$^-$ molecule. For atom indexing see Fig.~4a.}
\begin{tabular}{lcccc}
\hline
\hline
&\multicolumn{2}{c}{SnO$_2$}&\multicolumn{2}{c}{TiO$_2$}  \\
       & CA      & BP   & CA      & BP        \\ \hline
O$_{\rm I}$ -- M$_{\rm I}$ &   1.2\%  &   1.5\% &   2.5\%  &   2.7\%  \\
O$_{\rm IV}$ -- M$_{\rm II}$ &   2.2\%  &   3.4\% &   3.8\%  &   4.3\%  \\
O$_{\rm II}$ -- M$_{\rm II}$ &   1.1\%  &   1.1\% &   1.4\%  &   0.9\%  \\
O$_{\rm II}$ -- M$_{\rm I}$ &  -1.5\%  &  -1.4\% &  -1.7\%  &  -1.0\%  \\
O$_{\rm III}$ -- M$_{\rm I}$ &  -1.5\%  &  -1.9\% &  -2.7\%  &  -3.1\%  \\
&&&&\\
O$_{\rm I}$ -- H$_{\rm I}$ &  -0.8\%  &  -0.8\% &  -1.6\%  &  -1.5\%  \\
O$_{\rm A}$ -- H$_{\rm II}$ &  -2.2\%  &  -2.2\% &  -1.7\%  &  -2.0\%  \\
O$_{\rm A}$ -- M$_{\rm II}$ &  -5.5\%  &  -6.2\% &  -6.9\%  &  -7.0\%  \\
\hline
\hline
\end{tabular}
\end{table}

Figure~5a shows both total and local DOS
calculated for the hydroxylated (110) surface of SnO$_2$ and TiO$_2$.
Two oxygen sites were studied using the LDOS: the bridging
oxygen atom O$_{\rm I}$ and the oxygen of the adsorbed OH$^-$ group O$_{\rm
A}$.
Compared with clean surfaces, two main modifications are apparent:
\begin{figure}
\caption{Calculated valence band densities of states for
hydroxylated (110) surfaces of TiO$_2$ and SnO$_2$ in a) SD, b) UD
c) SM adsorption geometries. The dashed line
represents the LDOS on the bridging oxygen site O$_{\rm I}$, and the dotted
line
the LDOS on the adsorbed oxygen site O$_{\rm A}$.  For the sake
of presentation the local contributions have been scaled by a factor of 5.}
\end{figure}
\begin{itemize}
\item the contribution due to the bridging oxygen atom, situated at
the top of the valence band for clean surfaces (and giving an
extra feature above the O(2s) band)
has been pushed towards lower energies, and hybridizes
more strongly with the bulk O(2p) band.
On the other hand, formation of the O--H$^+$ bond gives rise to sharp
bonding 3$\sigma$ states
below the O(2p) band and 2$\sigma$ states below the O(2s) band.
\item the adsorbed hydroxyl group gives a contribution to
the valence O(2p) band due to its occupied 1$\pi$ state. It
lies above the surface VBM in the case of SnO$_2$~(110) and
within the surface VB for TiO$_2$~(110). On the other hand,
bonding 3$\sigma$ states can be seen as a narrow peak below
the O(2p) band on TiO$_2$~(110), whereas on SnO$_2$~(110) they
lie within the O(2p) band.
For both materials, the 2$\sigma$ peak lies within O(2s) band.
Separations of $2\sigma - 3\sigma$ and $3\sigma - 1\pi$ are respectively
10.3~eV and 6.2~eV for SnO$_2$ and 11.4~eV and 5.3~eV for TiO$_2$.

\end{itemize}
It is clear from the results that, especially for the adsorption energy,
gradient corrections make
a substantial difference. All the remaining calculations are performed
with the Becke-Perdew GGA scheme only.

\subsection{Unsymmetrical dissociative adsorption (UD)}
The SD geometry described in the previous section is
not the most stable one, and a small displacement of the adsorbed atoms
from their fully symmetric positions makes the system relax to a
configuration of lower energy. We find that this unsymmetrical
dissociative (UD) configuration gives
an adsorption energy of 1.39~eV per H$_2$O for
SnO$_2$~(110) and 1.08~eV per H$_2$O for TiO$_2$~(110),
so that the breaking of symmetry yields a stabilization of
well over 0.5~eV.

The nature of the new relaxed configuration is shown in Fig.~4b.
The bridging (O$_{\rm I}$) and adsorbed (O$_{\rm A}$) oxygens
approach each other, and the proton (H$_{\rm I}$) attached to bridging oxygen
tilts towards O$_{\rm A}$, so as to form a
hydrogen bond O$_{\rm A}$--H$_{\rm I}$ of
length 1.81~\AA\ on SnO$_2$~(110) and 1.80~\AA\ on TiO$_2$~(110). The
separation between O$_{\rm A}$ and O$_{\rm I}$
oxygens is 2.78 and 2.77~\AA\ for
the two materials. These changes
induce a dilation of the O$_{\rm I}$--H$_{\rm I}$ bond (by
5.2\% for SnO$_2$ and 4.4\% for TiO$_2$) and smaller
dilations of bonds between oxygens and surface cations.

Relaxation to the unsymmetrical configuration causes noticeable changes to
the DOS and LDOS described in the previous section. We note (see Fig.~5b)
a significant broadening of the isolated peak due to non-bonding
$1\pi$ states of adsorbed OH$^-$. In both materials, the $3\sigma$
levels are shifted upwards and hybridize more strongly with the
valence band. Similar changes can be observed in the LDOS of the
O$_{\rm I}$--H$_{\rm I}$ group, with a marked upward shift of 3$\sigma$ into
the O(2$p$) band. The separations of the $2\sigma - 3\sigma$ and $3\sigma -
1\pi$ peaks become respectively 11.7~eV and 4.6~eV for SnO$_2$ and
13.3~eV and 3.7~eV for TiO$_2$.

\subsection{Symmetrical molecular adsorption (SM)}
We have considered the case of symmetrical molecular (SM) adsorption
in which the water molecule bonds by its oxygen to the
surface 5-fold coordinated cation and the plane of the molecule is
the (001) plane (Fig.~4c).
We find that for both materials adsorption in this geometry is
energetically favorable, the adsorption energies being
0.78~eV per H$_2$O for SnO$_2$ and 0.82~eV per H$_2$O for TiO$_2$. These
are considerably larger than the adsorption energies for
the SD geometry. (Recall that we are comparing
energies calculated with the GGA.)

We find that molecular adsorption of H$_2$O causes only minor changes to the
surface structure. Relative to the relaxed clean surface, the 5-fold
coordinated cation moves out by 0.05~\AA (SnO$_2$) and 0.03~\AA (TiO$_2$). The
bridging oxygen moves out somewhat less and the atoms of
the first M$_2$O$_2$
atomic plane move inwards. The size of the latter effect is nearly negligible
for SnO$_2$~(110) but the displacements are as
large as 0.1~\AA\ on TiO$_2$~(110).
There is also a significant deformation of the adsorbed molecule:
even though the O--H bond lengths
remain unchanged, the angle between the bonds is increased
by 10\%. This is mainly due to modification of the water 3a$_1$
orbital, because of its contribution to bonding to the surface cation.
The O-cation bond between the water molecule and the surfaces
is about 10\% longer than the corresponding bond in the bulk
crystal.

Densities of states for molecularly adsorbed water are
presented in Fig.~5c.
Compared with the DOS of the clean surface, the features attributed to the
bridging oxygens (additional peak above O(2s) and O(2p) bands)
remain practically unchanged. The modification due to
adsorbed H$_2$O can be seen as an isolated narrow peak
below the O(2s) and O(2p) bands, as well as a two-peak
contribution to the O(2p) band. These can be attributed
to 2a$_1$, 1b$_2$, 3a$_1$ and 1b$_1$ molecular states
respectively. The energy separations between peaks are
about 11.5/2.9/3.4~eV for SnO$_2$ and 11.5/3.4/3.2~eV for TiO$_2$,
which are quite
close to the free molecule results (12.0/3.6/2.1~eV),
with the biggest modification, as expected, being the
downward shift and a small splitting (for SnO$_2$~(110)) of the 3a$_1$ peak.

\subsection{Surface dissociation}
Of the geometries we have examined, the two most stable
are the UD and the SM configurations, with the difference of
adsorption energies being
$\sim$~0.6~eV for SnO$_2$ and 0.3~eV for TiO$_2$. In both
materials, dissociative adsorption is favored. It thus seems
possible that water adsorbed molecularly on the
surface can spontanously dissociate.
Whether or not this occurs
will depend crucially on the existence and height of the
energy barrier that has to be overcome when the
hydrogen bond is created and when
the proton migrates along it.

To investigate this problem, we have calculated the total energy
of the system for a number of configurations along the probable
dissociation path. In practice, we have chosen the reaction coordinate
to be the horizontal (in the surface plane) separation
of the 5-fold coordinated cation and the migrating proton.
The positions of all other atoms, as well as the vertical position
of 5-fold coordinated cation and of H$_{\rm I}$ have been relaxed
for each value of the reaction coordinate.
Results on the dependence of the adsorption energy
on the M$_{\rm I}$--H$_{\rm I}$
horizontal separation for both materials are displayed in Fig.~6.
\begin{figure}
\caption{Dependence of adsorption energy on the reaction coordinate
(see text) for water dissociation on SnO$_2$~(110) and TiO$_2$~(110).}
\end{figure}
For both materials, we find almost no energy barrier which would need to
be overcome on the passage from SM to UD adsorption geometries.
It is possible that there may be a very shallow minimum at the SM
geometry, but this would not be of any practical significance.

\subsection{Adsorption in the low adsorbate density limit}
In order to get insight into the dependence of adsorption
characteristics on the adsorbate density we have considered
the half-coverage case where there is one H$_2$O in every two surface
unit cells. Because of limitations on computer time, we have chosen a
single adsorption geometry, namely the
UD one, which is energetically the most
favorable in the high density limit.

The practical calculations were performed with a supercell
twice as big as the previous one, keeping however
a slab thickness of three O - M$_2$O$_2$ - O units and a vacuum width
equivalent to two units. The clean surface
calculations were repeated for this supercell in order
to ensure cancellation of errors between the clean and hydroxylated slabs.
The positions of all slab and adsorbate atoms were relaxed to equilibrium.
We find that the adsorption energy is greater for the half-coverage case,
namely 1.63~eV per H$_2$O for SnO$_2$~(110).
Compared with full coverage case, the atomic structure of the surface and
adsorbate are only slightly modified.

\section{Discussion}
Our calculations on the energies of different relaxed configurations
indicate that dissociative adsorption gives
the largest adsorption energy and that the most stable configuration
is unsymmetrical. The adsorption energies at full
coverage are 1.39 and 1.08~eV per water molecule for SnO$_2$ and TiO$_2$,
increasing to 1.63 for SnO$_2$ at half coverage.
It is important to note that dissociative adsorption gives the
strongest binding, even though we are dealing with the perfect
surface and no defects are involved.
This appears to be an entirely
geometrical effect, since our results are very similar for SnO$_2$ and
TiO$_2$, in spite of their different electronic structure. In fact, our
entire set of results both for molecular and for dissociative
adsorption, and for the instability of the molecularly adsorbed state, is
remarkably similar for the two materials. Their similarity may perhaps
be understood by noting that they differ mainly in their {\em unoccupied}
electronic states.

In the TDS experiments referred to in the Introduction \cite{hug94,ger95},
the high temperature features at 435~K (SnO$_2$) and 375~K (TiO$_2$) were
attributed to desorption from the dissociated state. Adsorption energies
can be deduced from these desorption temperatures following the
analysis of Redhead~\cite{red62}, which expresses the desorption energy rate as
the
product of an effective vibrational frequency and a Boltzman factor
$\exp(-E_{\rm ads}/kT)$, where $E_{\rm ads}$ is the adsorption energy.
Assuming a frequency of 10$^{13}$ s$^{-1}$, we find that this
analysis gives adsorption energies of 1.1 and 1.0~eV for SnO$_2$ and TiO$_2$
respectively.
Our calculated values are thus in fair agreement with experiment,
though they are systematically too high. It is likely that this discrepancy
is related to the fact that our DFT calculations -- even with gradient
corrections -- underestimate the energy of the reaction H$_2$O~$\rightarrow$
OH$^-$~+ H$^+$ (see sec. 3.2). The fact that the adsorption energy for
SnO$_2$ is greater than that for TiO$_2$ is correctly given by our
calculations.

The relation of our results to the two lower temperature peaks in the
TDS spectra depends on how one interprets these peaks. The peak at the
lowest temperature appears to arise from desorption of molecular
H$_2$O when more than a monolayer is present, i.e. desorption
of H$_2$O bound to other H$_2$O. Our calculations clearly have nothing
to say about this. The intermediate peak at 300~K (SnO$_2$) and
275~K (TiO$_2$) was also attributed mainly to desorption from the
molecularly bound state. Our calculations clearly suggest that molecular
H$_2$O directly bound to the (110) surface is unstable with respect
to dissociation.
It is relevant to note that the semi-empirical calculations
of Bredow and Jug~\cite{bre95} on the isolated H$_2$O molecule
on the TiO$_2$ (110) surface found a barrier separating the
molecular and dissociated states, but the height of this
barrier was only $\sim$~0.2~eV.
There are then two possible interpretations of the
experiments: either the water desorbed at the intermediate temperature is
not bound directly to the oxide, but is perhaps bound to hydroxyl groups;
or there is some more stable molecularly adsorbed state which we have not
examined. Both of these seem quite plausible. The second interpretation
could be probed by making a more extensive search for stable molecularly
adsorbed configurations, perhaps using dynamical simulations, and we
hope to return to this.

Our calculations have predicted an appreciable increase of dissociative
adsorption energy with decreasing coverage. This is consistent with
experimental findings that the coverage of dissociatively adsorbed
water on SnO$_2$ and TiO$_2$~(110) is less than a monolayer, and can be
attributed to the effect of repulsion between hydroxyl groups. The same
effect has been found in semi-empirical quantum calculations on the
dissociative adsorption of H$_2$O on TiO$_2$~\cite{gon95,gon93}.

Calculated local densities of states for the oxygen atom $O_{\rm A}$
in the adsorbed OH$^-$ group (geometries SD and UD) or
for the adsorbed water molecule (geometry SM) strongly resemble
the spectra of the corresponding free molecules. The principal modification
concerns the molecular orbital directly involved in formation of the bond
with the surface, $3\sigma$ for OH$^-$ and $3a_1$ for H$_2$O. In most
cases modifications consist of a downward shift (relative to the
free molecule) accompanied by a small splitting, and are consistent with
UPS findings on hydroxylated TiO$_2$~(110)~\cite{kur89}.
It is also worth noticing, that for the more stable unsymmetrical
geometry the peak related to nonbonding $1\pi$ or $1b_1$ is considerably
broadened or split by the formation of a hydrogen bond and hybridization with
the surface O(2p) band.
On the other hand adsorption of the proton on the bridging oxygen introduces
a strong modification to its LDOS. This is clearest for the symmetrical
dissociative adsorption case, where the additional electrostatic field
of adsorbed H$^+$ causes a substantial downward shift of the
totality of LDOS(O$_{\rm I}$) compared to the clean surface.
In addition a distinct peak due to bonding O--H states appears below
O(2p) band. The overall structure of the LDOS on bridging oxygen
for dissociative adsorption is quite similar
to the free molecule spectrum, although for the unsymmetrical geometry
broadening and splitting
of  both features in the region of the O(2p) band become important.

Comparison of our results with experimental UPS spectra~\cite{kur89,ger95}
shows quite
good agreement. However it should be kept in mind that experimental spectra
taken at low temperatures tend to show molecularly adsorbed water which  may
be in higher adsorbed layers, rather than being bound directly to the surface.
This could explain the much closer resemblance
of measured spectra to those of free molecules.
On the other hand, the high-temperature spectra showing a two-peak structure
which can be assigned to OH$^-$ molecular orbitals relate well to our
LDOS for adsorbed oxygen. A surprising feature of the experimental results
is the total absence of signal from OH groups formed on bridging oxygens, which
according to our calculations should also give a characteristic,
two-peak structure.

\section{Conclusions}
We have studied molecular and dissociative adsorption of water on SnO$_2$~(110)
and SnO$_2$~(110) by an {\em ab initio}, density functional approach.
We found that inclusion of gradient corrections to the LDA noticeably
reduces the adsorption energies, a tendency which has been found already
for surface adsorption on metals.
For both materials, in the full coverage regime, we find that both molecular
and dissociative adsorption are energetically favorable. However,
dissociative adsorpton has a substantially greater adsorption energy.
Investigation of a possible reaction path for the surface dissociation of water
shows the absence of any energy barrier.
We found an increase of adsorption energy in the half-coverage
regime, which indicates the existence of an effective repulsive interaction
between adsorbed species and suggests that the most stable adsorption
configurations belong to the low coverage limit.
Calculated valence band DOS spectra for molecular and dissociative
adsorption geometries show structures similar to these of free
H$_2$O and OH$^-$ molecules. Bonding to the surface introduces
small shifts and broadening of peaks in the region of the surface valence band.

\begin{ack}
The work of JG is supported by EPSRC grant GR/J34842. The major
calculations were performed on the Intel iPSC/860 parallel computer
at Daresbury Laboratory, and we are grateful for a generous
allocation of time on the machine. Analysis of the results was
performed using local hardware funded by EPSRC grant GR/J36266.
Assistance from L.~N.~Kantorovich, J.~M.~Holender and
J.~A.~White is also acknowledged.

\end{ack}


\begin{thebibliography}{99}

\bibitem{gon95} J.~Goniakowski, C.~Noguera, Surf.~Sci.~{\bf 330} (1995) 337.
\bibitem{lan94} W.~Langel, M.~Parrinello, Phys.~Rev.~Lett.~{\bf 73} (1994) 504.
\bibitem{sca94} C.A.~Scamehorn, N.M.~Harrison, M.I.~McCarthy,
J.~Chem.~Phys.~{\bf 101} (1994) 1547.
\bibitem{tan89} J.~Tamaki, M.~Nagaishi, Y.~Teraoka, N.~Miura, N.~Yamazoe,
K.~Moriya, Y.~Nakamura, Surf.~Sci.~{\bf 221} (1989) 183.
\bibitem{fuj72} A.~Fujishima, K.~Honda, Nature~{\bf 283} (1972) 37.
\bibitem{hen77} V.E.~Henrich, G.~Dresselhaus, H.J.~Zeiger, Solid State
Commun.~{\bf 24} (1977) 623.
\bibitem{kur89} R.L.~Kurtz, R.~Stockbauer, T.E.~Madey, E.~Rom\'an,
J.L.~de~Segovia, Surf.~Sci.~{\bf 218} (1989) 178.
\bibitem{pan92} J.-M.~Pan, B.L.~Maschhoff, U.~Diebold, T.E.~Madey,
J.~Vac.~Sci.~Technol. A~{\bf 10} (1992) 2470.
\bibitem{mur91} C.A.~Muryn, P.J.~Hardman, J.J.~Crouch, G.N.~Raiker,
G.~Thornton, Surf.~Sci.~{\bf 251/252} (1991) 747.
\bibitem{hug94} M.B.~Hugenschmidt, L.~Gamble, C.T.~Campbell, Surf.~Sci.~{\bf
302} (1994) 329.
\bibitem{mun71} G.~Munera, F.S.~Stone, Disc.~Faraday~Soc.~{\bf 52} (1971) 205.
\bibitem{jon71} P.~Jones, J.A.~Hockey, Trans.~Faraday~Soc.~{\bf 67} (1971)
2669, 2679.
\bibitem{ger95} V.A.~Gercher, D.F.~Cox, Surf.~Sci.~{\bf 322} (1995) 177.
\bibitem{jay74} M.J.~Jaycock, J.C.R.~Waldsax, J.~Chem.~Soc. Faraday~Trans.
I~{\bf 70} (1974) 1501.
\bibitem{tsu83} M.~Tsukada, H.~Adachi, C.~Satoko, Prog.~Surf.~Sci.~{\bf 14}
(1983) 113.
\bibitem{gon93} J.~Goniakowski, S.~Bouette-Russo, C.~Noguera, Surf.~Sci.~{\bf
284} (1993) 315.
\bibitem{fah94} A.~Fahmi, C.~Minot, Surf.~Sci.~{\bf 304} (1994) 343.
\bibitem{bre95} T.~Bredow, K.~Jug, Surf.~Sci.~{\bf 327} (1995) 398.
\bibitem{vit92} A.~De~Vita, M.J.~Gillan, J.-S.~Lin, M.C.~Payne, I.~$\check{\rm
S}$tich, L.J.~Clarke, Phys.~Rev.~B~{\bf 46} (1992) 12964.
\bibitem{Ian92} I.~Manassidis, A.~De~Vita, J.-S.~Lin, M.J.~Gillan,
Europhys.~Lett.~{\bf 19} (1992) 605.
\bibitem{Ian93} I.~Manassidis, A.~De~Vita, M.J.~Gillan, Surf.~Sci.~Lett.~{\bf
285} (1993) L517.
\bibitem{ram94a} M.~Ramamoorthy, R.D.~King-Smith, D.~Vanderbilt,
Phys.~Rev.~B~{\bf 49} (1994) 7709.
\bibitem{ram94b} M.~Ramamoorthy, D.~Vanderbilt, R.D.~King-Smith,
Phys.~Rev.~B~{\bf 49} (1994) 16721.
\bibitem{IanYY} I.~Manassidis, M.J.~Gillan, Phys.~Rev.~B, submitted.
\bibitem{IanXX} I.~Manassidis, J.~Goniakowski, L.N.~Kantorovich, M.J.~Gillan,
accepted Surf.~Sci.
%\bibitem{Laa93} K.~Laasonen, M.~Sprik, M.~Parrinello, R.~Car,
%%J.~Chem.~Phys.~{\bf 99} (1993) 9080.
\bibitem{bec88} A.D.~Becke, Phys.~Rev.~A~{\bf 38} (1988) 3098.
\bibitem{per86} J.P.~Perdew, Phys.~Rev.~B~{\bf 33} (1986) 8822; {\bf 34} (1986)
7406(E).
\bibitem{hoh64} P.~Hohenberg, W.~Kohn, Phys.~Rev. {\bf 136} (1964) B864.
\bibitem{koh65} W.~Kohn, L.~J.~Sham, Phys.~Rev. {\bf 140} (1965) A1133.
\bibitem{jon89} R.~O.~Jones, O.~Gunnarsson, Rev.~Mod.~Phys. {\bf 61} (1989)
689.
\bibitem{gil91} M.~J.~Gillan, in {Proc. NATO ASI on Computer Simulation
in Material Science, Aussois, March 1991}, ed. M.~Mayer and V.~Pontikis,
p.~257 (Dordrecht, Kluwer, 1991).
\bibitem{pay92} M.C.~Payne, M.P.~Teter, D.C.~Allan, T.A.~Arias,
J.D.~Joannopoulos, Rev.~Mod.~Phys.~{\bf 64} (1992) 1045.
\bibitem{cla92} L.J.~Clarke, I.~$\check{\rm S}$tich, M.C.~Payne,
Comput.~Phys.~Commun.~{\bf 72} (1992) 14.
\bibitem{lan83} D.C.~Langreth, M.J.~Mehl, Phys.~Rev.~B~{\bf 20} (1983) 1809.
\bibitem{lan85} C.D.~Hu, D.C.~Langreth, Phys.~Scr.~{\bf 32} (1985) 391.
\bibitem{pw86} J.P.~Perdew, Y.~Wang, Phys.~Rev.~B~{\bf 33} (1986) 8800.
\bibitem{kut88} F.W.~Kutzler, G.S.~Painter, Phys.~Rev.~B~{\bf 37} (1988) 2850.
\bibitem{bos90} P.~Boschan, H.~Gollisch, Z.~Phys.~D~{\bf 17} (1990) 127.
\bibitem{mly91} P.~M\l ynarski, D.R.~Salahub, Phys.~Rev.~B~{\bf 43} (1991)
1399.
\bibitem{gar92} A.~Garc\'{\i}a, C.~Els\"asser, J.~Zhu, S.G.~Louie, M.L.~Cohen,
Phys.~Rev.~B~{\bf 46} (1992) 9829.
\bibitem{kon90} X.J.~Kong, C.T.~Chan, K.M.~Ho, Y.Y.~Ye, Phys.~Rev.~B~{\bf 42}
(1990) 9357.
\bibitem{ort92} G.~Ortiz, Phys.~Rev.~B~{\bf 45} (1992) 11328.
%\bibitem{zhu92} J.~Zhu, X.W.~Wang, S.G.~Louie, Phys.~Rev.~B~{\bf 45} (1992)
%%8887.
%\bibitem{hag93} J.~Haglund, Phys.~Rev.~B {\bf 47} (1993) 566.
\bibitem{whi94a} J.A.~White, D.M.~Bird, M.C.~Payne, I.~$\check{\rm S}$tich,
Phys.~Rev.~Lett.~{\bf 73} (1994) 1404.
\bibitem{hu94} P.~Hu, D.A.~King, S.~Crampin, M.H.~Lee, M.C.~Payne,
Chem.~Phys.~Lett. {\bf 230} (1994) 501.
\bibitem{phi94} P.H.T.~Philipsen, G.~Tevelde, E.J.~Baerends, Chem.~Phys.~Lett.
{\bf 226} (1994) 583.
\bibitem{gun94} K.~Gundersen, K.W.~Jacobsen, J.K.~N\o rskov, B.~Hammer,
Surf.~Sci. {\bf 304} (1994) 131.
\bibitem{unp} J.~Goniakowski, J.M.~Holender, L.N.~Kantorovich, M.J.~Gillan,
J.A.~White, Phys.~Rev. B, to be submitted
\bibitem{whi94} J.A.~White, D.M.~Bird, Phys.~Rev.~B~{\bf 50} (1994) 4954.
\bibitem{cep80} D.M.~Ceperley, B.J.~Alder, Phys.~Rev.~Lett.~{\bf 45} (1980)
566.
\bibitem{kle82} L.~Kleinman, D.M.~Bylander, Phys.~Rev.~Lett.,~{\bf 48} (1982)
1425.
\bibitem{lin93} J.-S.~Lin, A.~Qteish, M.C.~Payne, V.~Heine, Phys.~Rev.~B~{\bf
47} (1993) 4174.
\bibitem{jep71} O.~Jepsen, O.K.~Andersen, Solid State Comm.~{\bf 9} (1971)
1763.
\bibitem{leh72} G.~Lehmann, M.~Taut, Phys.~Stat.~Sol.~{\bf 54} (1972) 469.
\bibitem{mon76} H.J.~Monkhorst, J.D.~Pack, Phys.~Rev.~B~{\bf 13} (1976) 5188.
\bibitem{wyckoff} R.~Wyckoff, {\em Crystal Structures}, 2nd.~ed., vol.~1
(Interscience, New York, 1964).
\bibitem{les85} M.~Leslie, M.J.~Gillan, J.~Phys.~C: Solid State Phys.~{\bf 18}
(1985) 973.
\bibitem{eisenberg} D.~Eisenberg, W.~Kauzmann, {\em The Structure and
Properties of Water}, Oxford University Press, 1969.
\bibitem{herzberg} G.~Herzberg, {\em Spectra of Diatomic Molecules}, 2nd~ed.
D.~Van~Nostrand \& Company, Inc., Princeton, N.J., 1950.
\bibitem{bal78} R.E.~Ballard, {\em Photoelectron Spectroscopy and Molecular
Orbital Theory}, A.~Hilger~Ltd., Bristol~1978.
\bibitem{red62} P.A.~Redhead, Vacuum~{\bf 12} (1962) 203.
\end{thebibliography}
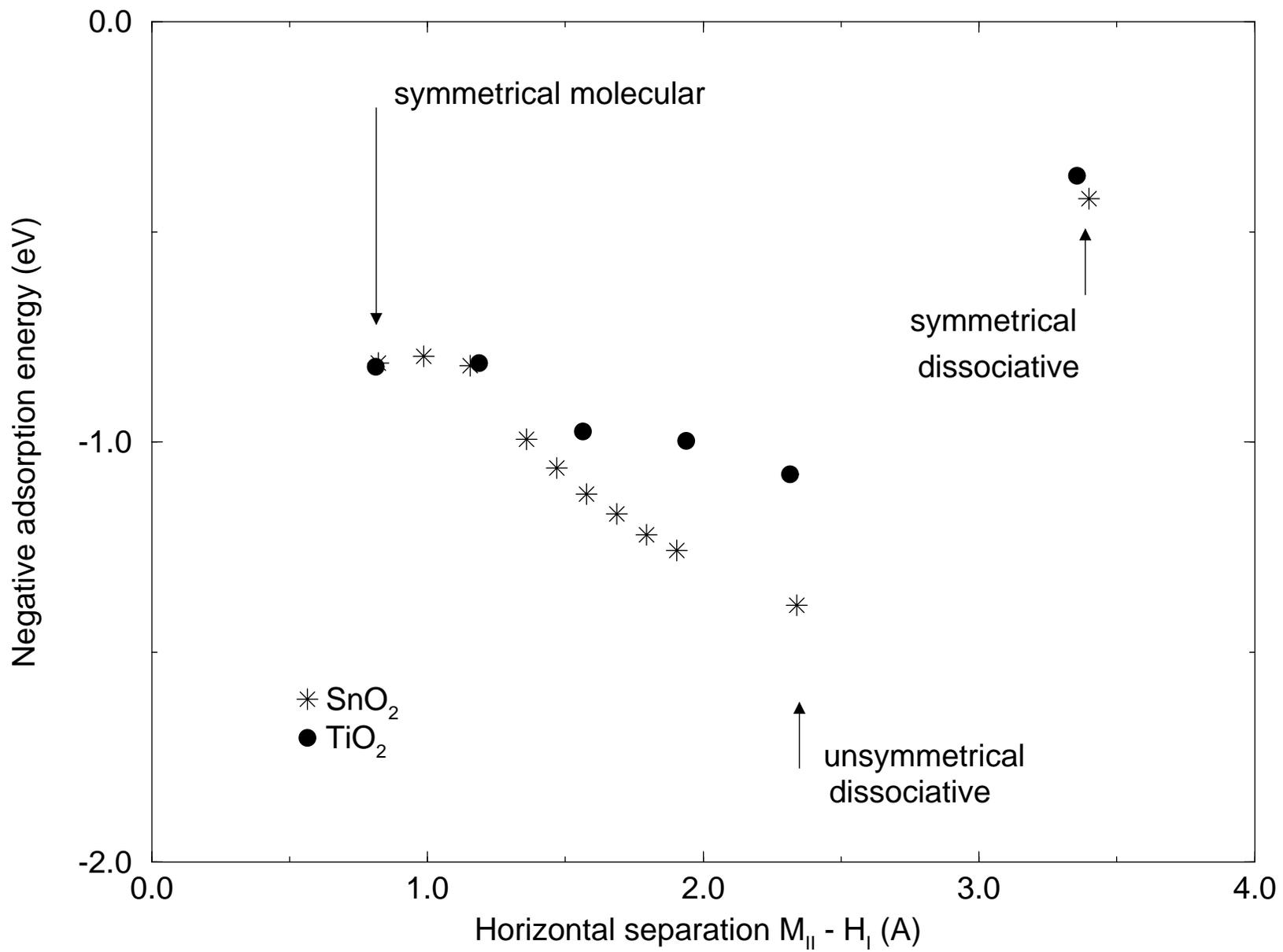
\end{document}